\documentclass[reprint,amsmath,amssymb,aps,prl]{revtex4-2}

\usepackage{graphicx}
\usepackage{dcolumn}
\usepackage{bm}
\usepackage{tikz}
\usepackage{hyperref}
\graphicspath{{plots/}}

\begin{document}

\preprint{SRFELO paper 1}

\title{Realizing A Hard X-Ray Storage Ring Free Electron Laser Oscillator at the APS-U}

\author{Emmanuel Aneke}
\email{eaneke@u.northwestern.edu}
\affiliation{Applied Physics Program, Northwestern University, Evanston, Illinois 60208, USA}

\author{Chris Jacobsen}
\email{c-jacobsen@northwestern.edu}
\affiliation{Department of Physics \&{} Astronomy, Northwestern University, Evanston, Illinois 60208, USA}

\author{Ryan Lindberg}
\email[Corresponding author: ]{lindberg@anl.gov}
\affiliation{Argonne National Laboratory, 9700 S Cass Avenue, Lemont, Illinois 60439, USA}

\author{Kent P. Wootton}
\email[Corresponding author: ]{kwootton@anl.gov}
\affiliation{Argonne National Laboratory, 9700 S Cass Avenue, Lemont, Illinois 60439, USA}


\date{\today}

\begin{abstract}
  We show that the APS-U could support a hard X-ray storage ring
  free electron laser oscillator, providing a promising avenue toward high repetition rate, narrow bandwidth coherent light sources. The results of our numerical simulations demonstrate that a transverse gradient undulator yields a ${\geq}5\%$ single-pass gain at 8.05~keV and 10~keV. We further identify a configuration at 5~keV that does not require a TGU but still exceeds a 5\%{} gain threshold despite the relatively short 5-meter long insertion device. All cases presented retain spectral purity on the order of meV and reach a steady-state output whose equilibrium is consistent with the Renieri Saturation Limit. We have calculated the 5~keV case to have an average brightness ${\sim}10^{26}$~photons/$(\text{s} \cdot \text{mm}^2 \cdot \text{mrad}^2 \cdot 0.1\% \,\text{BW})$, representing an increase in more than four orders of magnitude from the standard APS-U undulator. These results indicate that a storage ring free electron laser oscillator at multi-keV photon energies is feasible with nominal APS-U parameters and standard x-ray cavity optics.
\end{abstract}

\maketitle


The invention of the visible light laser~\cite{maiman_nature_1960} heralded broad impact across scientific research, medicine and technology~\cite{slusher_1999_rmp}. Lasing has been demonstrated at
increasingly higher photon energies, starting first in the EUV and soft x-ray range
with pumped media lasers \cite{suckewer_1990_science,matthews_1995_nimb} and with
high harmonic gain \cite{bartels_science_2002}. The development of free electron lasers based on particle accelerators \cite{madey_japplphys_1971} has been especially significant, leading to X-ray Free Electron Lasers (XFELs)~\cite{pellegrini_epjh_2012} at energies above 1 keV. Over the last two decades, high peak brightness linac-based XFEL amplifier facilities \cite{pellegrini_2020_natrevphys} have developed in parallel to high average brightness storage-ring-based light sources \cite{zhao_2010_rast, hettel_jsr_2014}.

We are interested in the intersection of these two design approaches. Specifically, XFEL oscillators (XFELOs) represent a narrow meV-bandwidth coherent X-ray source, which could be especially impactful to X-ray techniques \cite{adams_workshop_2019} including coherent X-ray nanoimaging \cite{du_jac_2021} and non-resonant inelastic X-ray scattering \cite{wang_matradext_2020}. An XFELO could also be used for high-gain harmonic generation (HGHG) seeding of an FEL amplifier operating at higher photon energies $({\sim}40{-}100~\textrm{keV})$ \cite{marie_2019}.

XFELOs based on linacs and energy-recovery linacs have previously been proposed, incorporating ${>}20$~m-length undulators \cite{kim_2008_prl, dai_2012}. Experimental studies of XFELOs based upon MHz-repetition rate linacs have been performed at existing XFEL facilities \cite{white_2024, rauer_2026}, similarly incorporating long (${>}20$~m) undulators.

An attractive alternative is to use a storage ring as an electron beam source for a storage ring free-electron laser oscillator (SRFELO). Numerous experimental configurations of SRFELOs have been realized in the range of visible to vacuum-ultraviolet \cite{billardon_first_1983, superaco, niji} including lasing at a photon energy of 7.35~eV~\cite{wu_2021}. However, extending this approach further into X-ray photon energies has proven challenging. The first proposal for an SRFELO at $4{-}6$~keV photon energy identified its feasibility \cite{colella_1984_optcommun} using accelerator parameters not yet achieved, and furthermore no practical cavity mirror parameters existed at the time. 
In electron storage rings, the electron beam exhibits a larger energy spread and transverse emittance than in linac-based sources. This leads to a rapid degradation of the FEL resonance condition as the number of undulator periods increase. At shorter wavelengths, maintaining appreciable single-pass gain therefore requires progressively smaller
emittance and tighter energy spread to preserve longitudinal coherence
and transverse mode overlap over the interaction length. This constraint is particularly severe in the hard X-ray regime, where the short radiation wavelength worsens the three-dimensional gain reduction effects. Previous SRFELO geometries proposed at X-ray photon energies employed an undulator of typically ${\sim}$20--100~m length in a dedicated bypass line \cite{fisher_nima_1992, barletta_nima_2010, cai_2013_srn, lindberg_2013, agapov_nima_2015, agapov_ipac_2018, lee_2019, li_2023, hansen_2023_lund, yu_2024}.

\begin{figure}[htbp!]
    \centering
    \includegraphics[width=\columnwidth]{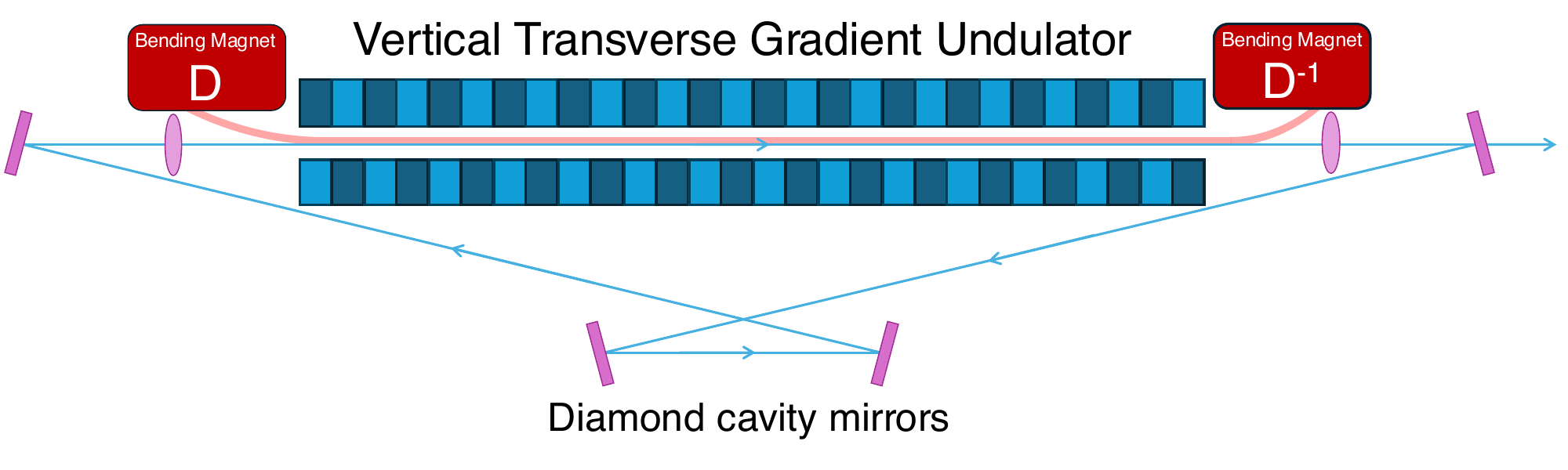}
    \caption{Layout of a TGU-based X-ray SRFELO.}
    \label{fig:layout}
\end{figure}

Contrasting with previous work, in this letter we demonstrate that a hard X-ray ($5{-}10~\textrm{keV}$) SRFELO is feasible within a standard insertion device straight ($5$~m length) of the Advanced Photon Source-Upgrade (APS-U) storage ring (Fig.~\ref{fig:layout}). Recent measurements of the emittance at an APS-U soft X-ray beamline \cite{aneke_sri2024, aneke_ipac2025} inspired us to investigate how the extremely bright beam might drive a next-generation light source. 
We found that both a planar undulator and a Transverse Gradient Undulator (TGU) \cite{first_tgu} can yield sufficient single-pass gain for a hard X-ray oscillator. We establish the viability of these configurations through cross-comparison of independent multi-pass simulation codes, and ultimately confirm in time-dependent tracking that the 5~keV planar baseline achieves a stable equilibrium and steady-state output.

One of the primary challenges in realizing a SRFELO is the relatively large energy spread of the stored electron bunch.  In a standard undulator of period $\lambda_u$, the resonant photon wavelength $\lambda$ is related to the particle Lorentz factor $\gamma$ according to
\begin{align}
\lambda = \frac{\lambda_u}{2\gamma^2}\left(1+\frac{K^2}{2}\right),
\end{align}
where $K$ is the dimensionless deflection parameter that is related to the peak magnetic field $B_0$ through the empirical formula $K \approx 0.934\lambda_u[\text{cm}]B_0[\text{T}]$.  Low-gain FELs amplify radiation about the resonant wavelength within a normalized bandwidth $\Delta\lambda/\lambda \sim 1/\pi N_u$, where $N_u$ is the number of undulator periods.  In storage rings, the energy spread $\sigma_\gamma/\gamma\sim 10^{-3}$ typically induces a spread in the resonant $\lambda$ larger than the FEL bandwidth, and FEL gain is significantly suppressed.

A TGU can help mitigate the effect of a large energy spread by first introducing a transverse magnetic field gradient across the poles leading to $K \rightarrow K_0(1+\alpha x)$. Next, one adds dispersion to the electron beam such that $x$-$\gamma$ correlation locally matches the transverse variation in the resonance due to the magnetic field gradient.  TGUs were previously considered to mitigate the energy spread effect for beams from both both laser wakefield accelerators \cite{huang_2012} and storage rings \cite{lindberg_2013}.

We have found that operating an SRFELO at the APS-U requires a TGU to overcome cavity losses for X-rays energies close to 10 keV.  Nevertheless, we also demonstrate that the baseline planar undulator case can generate 5 keV X-rays in a SRFELO.  This somewhat surprising result is due to a few factors. First, the 5-m straight section provides room for a device of $N_u \sim 200$ periods, whose corresponding FEL bandwidth $\Delta\lambda/\lambda \sim 10^{-3}$ is approximately the normalized energy spread, such that the gain degradation is modest \cite{lindberg_book}. Second, the equilibrium emittance of the APS-U is nearly diffraction limited at 5 keV, such that the FEL gain is not strongly impacted by the non-zero electron beam size and divergence.  Hence, the gain is only modestly reduced at 5 keV, and an undulator with $N_u \sim 200$ is sufficient for FEL lasing.


To establish a verified baseline for our SRFELO models, initial simulations were conducted using a custom Python-based multipass tracking framework  adapted from Ref.~\cite{li_2023}. This wrapper orchestrates a sequential, turn-by-turn simulation loop by coupling two highly established, independent codes: Genesis 1.3 \cite{reiche_1999_nima} for the single-pass FEL amplification within the insertion device, and ELEGANT \cite{elegant} for the six-dimensional electron beam transport through the storage ring lattice. During each simulated pass, Genesis calculates the radiation field growth, after which ELEGANT propagates the degraded electron bunch through the ring to accurately capture the evolution of the equilibrium emittance and energy spread. Between these discrete tracking steps, Python modules manage optical cavity physics. In this framework, the cavity was simplified to calculate gain and reflectivity by treating the optical layout as a sequence of discrete optical elements that contribute to the overall round-trip cavity losses. Because Genesis and ELEGANT are extensively benchmarked across the accelerator community, this sequential wrapper provided a highly reliable physical baseline.

FEL1X is an internal time-dependent simulation code initially designed to model XFELOs \cite{fel1x}. Now, the code is capable of simulating turn-by-turn dynamics of storage ring free-electron lasers. The continuous damping and diffusion/quantum excitation caused by by the emission of synchrotron radiation in the ring is modeled using a Fokker-Planck equation.
The FEL employs a standard method of discretizing the electron bunch into longitudinal bins that are populated by representative macroparticles \cite{lindberg_book}.
These macroparticles exchange energy with the field through the FEL interaction. This results in microbunching and FEL gain over a single pass, and spectral narrowing over many passes through the cavity.
Once we confirmed an alignment between the results of the Genesis wrapper and FEL1X, we transitioned the primary simulation effort to the fully self-consistent FEL1X code to capture the complete time-dependent dynamics.

To determine the optimal operating regimes, the study was anchored around three distinct configurations: a baseline planar undulator operating at 5~keV, and two TGU cases at 8.05~keV and 10~keV. Optimal undulator parameters were determined assuming out-of-vacuum devices that fit within the present APS-U straight section.  
The theoretical performance of these  
configurations was then evaluated analytically to maximize the small-signal gain without a TGU. For the top-performing candidates, parameter scans were then executed to identify the exact TGU factor that maximized the single-pass amplification. With the undulator and detuning parameters locked, the optical cavity geometry was next. Although asymmetric modes experience slightly larger gain, we assumed an axially symmetric optical mode with identical Rayleigh ranges in each plane for the cross correlation studies due to constraints in Genesis1.3. Finally, these optimized parameters were simulated using our full multi-pass tracking framework and cross-referenced against the self-consistent FEL1X code. Accelerator parameters used in simulation are summarized in Table~\ref{tab:parameters_accel}~\cite{fornek_2019}.

\begin{table}[htbp!]
    \centering
    \caption{Electron Beam and Accelerator Parameters \cite{fornek_2019}}
    \label{tab:parameters_accel}
    \begin{ruledtabular}
    \begin{tabular}{lcc}
        \textbf{Parameter} & \textbf{Symbol} & \textbf{Value} \\
        \hline
        Beam energy & $E$ & 6\,GeV \\
        Energy spread & $\sigma_\eta$ & 0.135\%\\
        Horizontal emittance & $\varepsilon_x$ & 42.0\,pm\,rad \\
        Vertical emittance & $\varepsilon_y$ & 4.2\,pm\,rad \\
        Bunch charge & $q_e$ & 3.6\,nC\\
        Horizontal beta function & $\beta_x^*$ & 5.2\,m\\
        Vertical beta function & $\beta_y^*$ & 2.4\,m\\
        Energy loss per turn & $U_0$            &  2.87\,MeV  \\
        Stored beam current  & $I$  & 50\,mA\\    
        Beam peak current  &  $I_\textrm{peak}$ & 72\,A\\ 
    \end{tabular}
    \end{ruledtabular}
\end{table}

The feasibility of planar and TGU insertion devices needed for the three cases is presented in Supplemental Material \cite{suppMaterial} (see also references \cite{casalbuoni_j_phys_conf_2025, kesgin_ieee_2021, bernhard_prab_2016, benabderrahmane_prab_2017} therein). The vertical dispersion selected for the 8.05 and 10~keV cases was $D=10.4$~mm. Insertion device parameters for the three oscillator cases simulated are summarized in Table~\ref{tab:parameters}.

\begin{table}[htbp!]
    \centering
    \caption{Insertion Device Parameters}
    \label{tab:parameters}
    \begin{ruledtabular}
    \begin{tabular}{lcccc}
        \textbf{Parameter} & \textbf{Symbol} & \textbf{5 keV} & \textbf{8.05 keV} & \textbf{10 keV}\\
        \hline
        Undulator period & $\lambda_u$ & 1.94\,cm & 1.7\,cm& 1.7\,cm \\
        Undulator length & $L_u$ & 4.998\,m & 5.0\,m& 5.0\,m \\
        Deflection parameter & $K_0$ & 2.25& 1.73& 1.42 \\
        Transverse gradient &$\alpha$ &0\,m$^{-1}$ & 213\,m$^{-1}$ & 254\,m$^{-1}$ \\
        TGU factor & $\Gamma$ & 0 & 3 & 3 \\
    \end{tabular}
    \end{ruledtabular}
\end{table}


The viability of hard X-ray optical cavities has recently been demonstrated~\cite{margraf_2023}. A detailed description of the Bragg-crystal optics model, including the dynamical diffraction treatment, complex reflectivity generation, and phase-flattening procedure used to remove crystal-induced delay, is provided in Supplemental Material \cite{suppMaterial2} (see also reference \cite{yuri_2010} therein).



\begin{figure}[htbp!]
\includegraphics[width=\columnwidth]{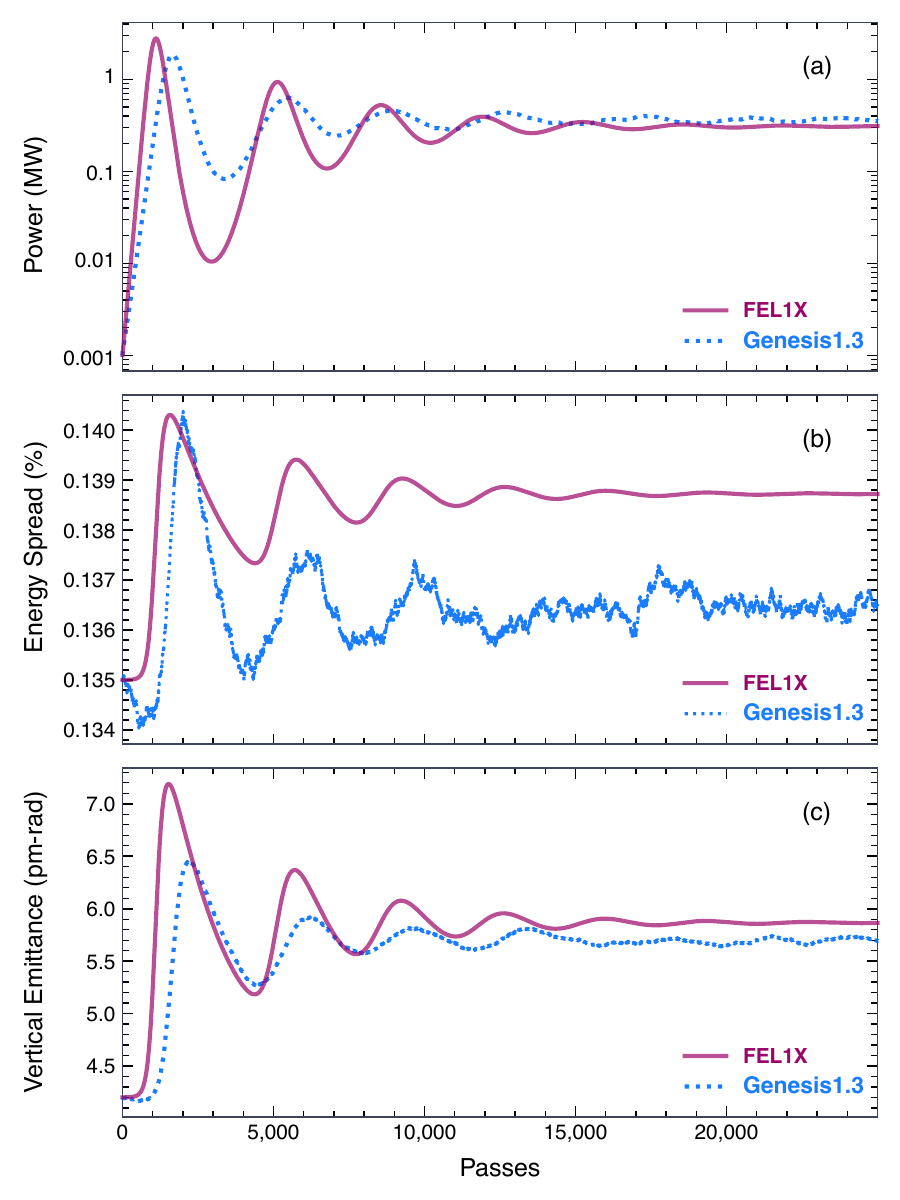}
\caption{\label{fig:10keV_stacked} Simulation of 10~keV SRFELO based on parameters in Tables~\ref{tab:parameters_accel} and \ref{tab:parameters}. (a) Simulation of laser pulse power in the cavity, as a function of passes in the optical cavity. For both codes, the power increases to an equilibrium state over ${\sim}$25000 passes of the optical cavity. (b) Simulation of electron beam energy spread. (c) Simulation of electron beam vertical emittance.}
\end{figure}

We demonstrate saturation of the 10~keV oscillator over ${\sim}$25000 passes, with the intracavity power, energy spread, and vertical emittance plotted as a function of passes in Fig.~\ref{fig:10keV_stacked}.  For this we assume an initial 1~kW of laser power, and the observed emittance growth occurs when the energy spread induced in the FEL is transferred to the vertical plane when the dispersion is removed after the undulator.

\begin{figure}[htbp!]
\includegraphics[width=\columnwidth]{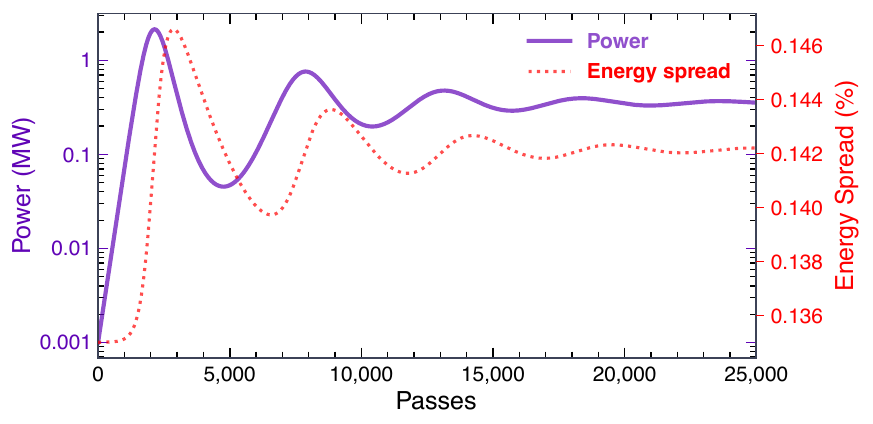}
\caption{\label{fig:5keV_singleSlice} Simulation in FEL1X of laser pulse power in the cavity and electron beam energy spread, as a function of passes in the oscillator cavity at 5~keV. Both the power and energy spread increases to an equilibrium state over ${\sim}$25000 passes of the optical cavity.  A few percent of the cavity power can be out-coupled for users.}
\end{figure}

To simplify this initial study to the case of a single x-ray pulse stored in the cavity, we considered the case of 50 mA stored beam current as indicated in Table~\ref{tab:parameters_accel}.
Figure~\ref{fig:5keV_singleSlice} plots the laser power and electron beam energy spread for the 5 keV oscillator that has no TGU, no dispersion, and therefore no emittance growth. Values for each of the 5, 8 and 10~keV oscillator laser and electron beam properties at equilibrium are summarized in Table~\ref{tab:results}.

\begin{table}[htbp!]
    \centering
    \caption{Time-Independent Properties at Equilibrium}
    \label{tab:results}
    \begin{ruledtabular}
    \begin{tabular}{lccc}
        \textbf{Parameter}   & \textbf{FEL1X} & \textbf{Genesis1.3} & \textbf{Units}\\
        \hline

        \textbf{5 keV}       &       & \\
        Gain & $5.5$      & $5.3$        &\% \\
        Power  & $0.36$     & $0.80$        &MW \\
        Energy spread  & $0.1422$     & $0.1447$       &\%  \\
        Vertical emittance & $4.20$     & $4.03$        &pm rad \\
        \textbf{8 keV}       &       & \\
        Gain & $7.1$      & $6.5$        &\% \\
        Power  & $0.970$     & $0.660$        &MW \\
        Energy spread & $0.1476$     & $0.1408$       &\%  \\
        Vertical emittance & $10.109$     & $8.203$        &pm rad \\
        \textbf{10 keV}       &       & \\
        Gain & $5.1$      & $4.6$        &\% \\
        Power  & $0.309$     & $0.364$        &MW \\
        Energy spread  & $0.1387$     & $0.1365$       &\%  \\
        Vertical emittance & $5.867$     & $5.718$        &pm rad \\
    \end{tabular}
    \end{ruledtabular}
\end{table}

\begin{figure}[htbp!]
\includegraphics[width=.95\columnwidth]{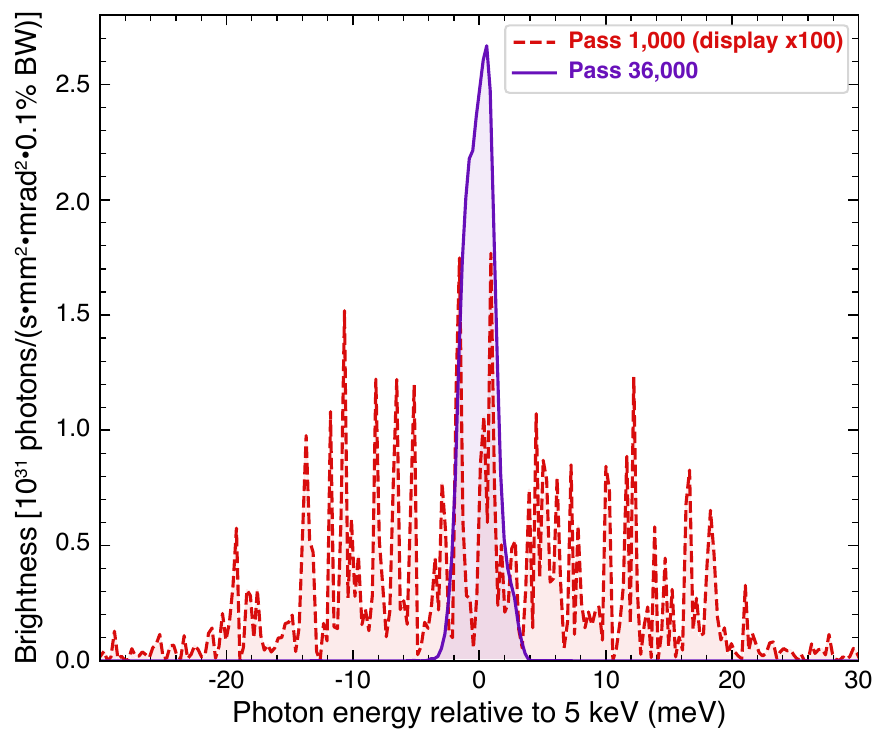}
\caption{\label{fig:5keVspectrum} Peak intracavity brightness of the 5~keV SRFELO simulated using FEL1X and the parameters in Tables~\ref{tab:parameters_accel} and \ref{tab:parameters}.}
\end{figure}

The time-dependent FEL1X simulations were propagated until the intracavity pulse energy and spectral distribution settled into a stable, quasi-steady state. Within this equilibrium regime, the transient fluctuations characteristic of the initial amplification phase are sufficiently damped by the storage ring lattice, allowing for a reliable extraction of the saturated output spectrum. Figure~\ref{fig:5keVspectrum} shows an example of a single-spiked spectrum at 5~keV, whose rms is ${\sim}1.5$~meV.  In the time domain the pulse duration is ${\sim}1.5$~ps.  Future efforts will study how the precise temporal and spectral profiles depend upon the FEL gain, cavity length detuning, and Bragg crystal parameters.  Nevertheless, the output of Fig.~\ref{fig:5keVspectrum} is calculated to have an average brightness ${\sim}10^{26}$~photons/$(\text{s} \cdot \text{mm}^2 \cdot \text{mrad}^2 \cdot 0.1\% \,\text{BW})$, representing an increase in more than four orders of magnitude from the standard APS-U undulator. This is brighter than the average brightness of LCLS-II-HE \cite{raubenheimer_2018_FLS}. Note that the difference between average and peak and brightness comes from the ratio of the pulse length to the pulse spacing.

Historically, achieving a steady-state, continuous-wave emission profile has proven complex, often resulting in periodic, spiked lasing behaviors \cite{li_2023, superaco}. In contrast, our simulations of the 5-meter configurations demonstrate a steady-state output. We suspect that this stable lasing is due to several factors.  First, the electron bunch length is much longer than the radiation pulse in equilibrium, so temporal overlap between the two varies little from pass to pass.  Second, the crystal bandwidth is much narrower than the natural FEL bandwidth $\sim 1/4N_u$, which effectively filters out frequency components that might drive the sideband/trapped particle instability \cite{goldstein_1985,colson_1986}.  Third, the equilibrium power is well below the maximum allowed by the Renieri saturation limit, which states that the average laser power of a storage ring FEL will not exceed a fraction of the total power lost in the whole machine via synchrotron radiation \cite{dattoli_1980, elleaume_macro-temporal_1984, dattoli_2019}.  Physically, the Renieri limit arises because the energy spread growth driven by the FEL-induced heating must be continuously balanced by the beam damping via synchrotron radiation cooling. While normally stated as a bound on the FEL power, if we consider the x-ray energy generated during one pass $U_{\textrm{FEL}}$, then the Renieri limit states that
\cite{dattoli_2019}:
\begin{equation}
U_{\textrm{FEL}} \le \frac{1}{4N_u}\frac{q_e}{e} U_0,    \label{eq:Renieri}
\end{equation}
where $U_0$ is the energy loss per turn in the ring and $q_e/e$ is the number of particles per bunch. For the APS-U parameters of Table~\ref{tab:parameters_accel}, the Renieri limit is $\sim 10 \;\mu\text{J}$ for $N_u=250$, which is slightly more than the steady-state stored energy in the 5-keV baseline, and considerably more than the $\sim 0.4 \;\mu\text{J}$ per pass that we observed to be generated in the FEL.


The SRFELO geometries presented in this letter are directly applicable to APS-U and the High Energy Photon Source with 5\,m length insertion device length and swap-out injection~\cite{jiao_aapps_2026, qin_ipac_2026a}. We reasonably expect that this could also be applied at other light source facilities including SPring-8-II \cite{tanaka_2026} and PETRA-IV \cite{li_2023}. With multiple storage ring projects currently in a design phase, it is possible multiple SRFELO user facilities could be pursued worldwide~\cite{shin_aapps_2021}. We have not yet explored possibilities for 3-GeV-class storage rings. Regarding tunability of the SRFELO, both the 8.05~keV and 10~keV TGU operating regimes can be driven by a single insertion device by tuning the magnetic gap and dispersive kick as outlined in Supplemental Material \cite{suppMaterial3}. In the present work, we have investigated only a single SRFELO in a machine. As future work, we intend to study whether a single ring may support several simultaneous and independent SRFELOs.


In conclusion, we have demonstrated through self-consistent turn-by-turn simulations that by exploiting the near-diffraction-limited emittance of fourth-generation storage rings, a conventional planar undulator provides sufficient gain to sustain oscillation at 5~keV. This directly contrasts with previous proposals in literature, which were predicated upon much longer undulator lengths, and typically proposed to be implemented within a dedicated bypass line. Moreover, our simulations indicate an approximately steady-state output consistent with the Renieri Saturation limit. This again contrasts with previous studies that either targeted a ``pulsed mode'' operation that periodically produced X-ray powers well above the Renieri limit, and differs from the family of longer wavelength SRFELOs that exhibited an unsteady, chaotic-appearing output. Furthermore, incorporating a TGU allows this compact footprint to support steady-state lasing at 8.05~keV and 10~keV by effectively mitigating energy spread degradation. These results establish a practical pathway toward realizing high-repetition-rate coherent X-ray sources using existing storage ring infrastructure.

\begin{acknowledgments}
This research (KW, RL, EA) used resources of the Advanced Photon Source, a U.S. Department of Energy (DOE) Office of
Science User Facility and is based on work supported by Laboratory Directed Research and Development (LDRD)
funding from Argonne National Laboratory, provided by the Director, Office of Science, of the U.S. DOE under
Contract No. DE-AC02-06CH11357.  This material also includes work supported (CJ, EA) by the Air Force Office of Scientific Research under award number FA9550-23-1-0284.
Any opinions, findings, and conclusions or recommendations expressed in this material are those of the author(s) and do not necessarily reflect the views of the United States Air Force.
\end{acknowledgments}

\bibliography{bibtex/references.bib}

@Preamble{
"\providecommand{\noopsort}[1]{}"
# "\providecommand{\singleletter}[1]{#1}%"
}

@article{bartels_science_2002,
	author = {Bartels, Randy A and Paul, Ariel and Green, Hans and Kapteyn, Henry C and Murnane, Margaret M and Backus, Sterling and Christov, Ivan P and Liu, Yanwei and Attwood, David T and Jacobsen, Chris},
	journal = {Science},
	month = jul,
	number = {5580},
	pages = {376--378},
	title = {Generation of spatially coherent light at extreme ultraviolet wavelengths},
	volume = {297},
	year = {2002},
    doi = {10.1126/science.1071718}
    }

@article{hettel_jsr_2014,
	author = {Hettel, Robert},
	journal = {J. Synchrotron Radiat.},
	month = sep,
	number = {5},
	pages = {843--855},
	title = {{DLSR} design and plans: an international overview},
	volume = {21},
	year = {2014},
    doi = {10.1107/S1600577514011515}}

@article{pellegrini_epjh_2012,
	author = {Pellegrini, C},
	journal = {Eur. Phys. J. H},
	doi = {10.1140/epjh/e2012-20064-5},
	month = jun,
	number = {5},
	pages = {659--708},
	title = {The history of {X-ray} free-electron lasers},
	volume = {37},
	year = {2012}}

@article{du_jac_2021,
	author = {Du, Ming and Di, Zichao (Wendy) and G{\"u}rsoy, Do{\u{g}}a and Xian, R. Patrick and Kozorovitskiy, Yevgenia and Jacobsen, Chris},
	journal = {J. Appl. Crystallogr.},
	month = feb,
	pages = {386--401},
	title = {{Upscaling X-ray nanoimaging to macroscopic specimens}},
	volume = {54},
	year = {2021},
    doi = {10.1107/S1600576721000194}
    }

@article{li_2023,
  title = {Transverse gradient undulator in a storage ring x-ray free electron laser oscillator},
  author = {Li, Yuanshen and Lindberg, Ryan and Kim, Kwang-Je},
  journal = {Phys. Rev. Accel. Beams},
  volume = {26},
  issue = {3},
  pages = {030702},
  numpages = {16},
  year = {2023},
  month = {Mar},
  publisher = {American Physical Society},
  doi = {10.1103/PhysRevAccelBeams.26.030702},
  url = {https://link.aps.org/doi/10.1103/PhysRevAccelBeams.26.030702}
}

@inproceedings{lindberg_2013,
    author = {R. R. Lindberg and K.-J. Kim and Y. Cai and Y. Ding and Z. Huang},
    title = {Transverse Gradient Undulators for a Storage Ring X-ray {FEL} Oscillator},
    booktitle = {Proc. FEL'13},
    year = 2013,
    pages = {740--748},
    paper = {THOBNO02},
    venue = {New York, NY, USA, Aug. 2013},
    publisher = {JACoW Publishing, Geneva, Switzerland},
    url = {https://jacow.org/FEL2013/papers/THOBNO02.pdf}
}

@book{lindberg_book, 
place={Cambridge}, 
title={Synchrotron Radiation and Free-Electron Lasers: Principles of Coherent X-Ray Generation}, 
publisher={Cambridge University Press}, 
author={Kim, Kwang-Je and Huang, Zhirong and Lindberg, Ryan}, 
year={2017}}

@article{wu_2021,
    author = {Wu, Y. K. and Mikhailov, S. and Yan, J. and Wallace, P. and Popov, V. and Pentico, M. and Swift, G. and Ahmed, M. W. and Kochanneck, L. and Ehlers, H. and Jensen, L. O.},
    title = {Lasing below 170 nm using an oscillator {FEL}},
    journal = {J. Appl. Phys.},
    volume = {130},
    number = {18},
    pages = {183101},
    year = {2021},
    month = {11},
    issn = {0021-8979},
    doi = {10.1063/5.0064942},
    url = {https://doi.org/10.1063/5.0064942}
}

@article{superaco,
    doi = {10.1209/0295-5075/21/9/006},
    url = {https://dx.doi.org/10.1209/0295-5075/21/9/006},
    year = {1993},
    month = {mar},
    publisher = {},
    volume = {21},
    number = {9},
    pages = {909},
    author = {M. E. Couprie and D. Garzella and M. Billardon},
    title = {Operation of the {Super-ACO Free-Electron Laser} in the {UV} Range At 800\,{MeV}},
    journal = {Europhys. Lett.}
}

@article{billardon_first_1983,
	title = {First {Operation} of a {Storage}-{Ring} {Free}-{Electron} {Laser}},
	volume = {51},
	issn = {0031-9007},
	url = {https://link.aps.org/doi/10.1103/PhysRevLett.51.1652},
	doi = {10.1103/PhysRevLett.51.1652},
	number = {18},
	urldate = {2024-11-21},
	journal = {Phys. Rev. Lett.},
	author = {Billardon, M. and Elleaume, P. and Ortega, J. M. and Bazin, C. and Bergher, M. and Velghe, M. and Petroff, Y. and Deacon, D. A. G. and Robinson, K. E. and Madey, J. M. J.},
	month = oct,
	year = {1983},
	pages = {1652--1655},
}

@article{niji,
title = {First lasing of the {NIJI-IV} storage-ring free-electron laser},
journal = {Nucl. Instrum. Methods Phys. Res., Sect. A},
volume = {331},
number = {1},
pages = {27-33},
year = {1993},
issn = {0168-9002},
doi = {10.1016/0168-9002(93)90008-6},
url = {https://doi.org/10.1016/0168-9002(93)90008-6},
author = {T. Yamazaki and K. Yamada and S. Sugiyama and H. Ohgaki and N. Sei and T. Mikado and T. Noguchi and M. Chiwaki and R. Suzuki and M. Kawai and M. Yokoyama and K. Owaki and S. Hamada and K. Aizawa and Y. Oku and A. Iwata and M. Yoshiwa}
}

@inproceedings{white_2024,
    author = {M. White and others},
    title = {{CBXFEL design, production, and installation status}},
    booktitle = {Proc. LINAC'24},
    pages = {758--761},
    paper = {THPB060},
    venue = {Chicago, IL, USA},
    series = {Linear Accelerator Conference},
    number = {32},
    publisher = {JACoW Publishing, Geneva, Switzerland},
    month = {8},
    year = {2024},
    issn = {2226-0366},
    isbn = {978-3-95-450219-6},
    doi = {10.18429/JACoW-LINAC2024-THPB060}
}

@article{rauer_2026,
  author    = {Rauer, Patrick and others},
  title     = {Lasing of a cavity-based {X}-ray source},
  journal   = {Nature (London)},
  year      = {2026},
  volume    = {650},
  number    = {8100},
  pages     = {93--96},
  doi       = {10.1038/s41586-025-10025-x},
  url       = {https://doi.org/10.1038/s41586-025-10025-x},
  issn      = {1476-4687}
}

@article{lee_2019,
    author = {Lee, Tae-Yeon},
    title = {Storage ring based x-ray {FEL} oscillator},
    journal = {AIP Conf. Proc.},
    volume = {2054},
    number = {1},
    pages = {030021},
    year = {2019},
    month = {01},
    issn = {0094-243X},
    doi = {10.1063/1.5084584},
    url = {https://doi.org/10.1063/1.5084584}
}

@article{dai_2012,
  title = {Proposal for an X-Ray Free Electron Laser Oscillator with Intermediate Energy Electron Beam},
  author = {Dai, Jinhua and Deng, Haixiao and Dai, Zhimin},
  journal = {Phys. Rev. Lett.},
  volume = {108},
  issue = {3},
  pages = {034802},
  numpages = {5},
  year = {2012},
  month = {Jan},
  publisher = {American Physical Society},
  doi = {10.1103/PhysRevLett.108.034802},
  url = {https://link.aps.org/doi/10.1103/PhysRevLett.108.034802}
}

@article{yu_2024,
  title = {{3D} small-gain formula allowing strong focusing and harmonic lasing for a ring-based x-ray free electron laser oscillator},
  author = {Yu, Li Hua and Smaluk, Victor and Shaftan, Timur and Tiwari, Ganesh and Yang, Xi},
  journal = {Phys. Rev. Accel. Beams},
  volume = {27},
  issue = {6},
  pages = {060702},
  numpages = {17},
  year = {2024},
  month = {Jun},
  publisher = {American Physical Society},
  doi = {10.1103/PhysRevAccelBeams.27.060702},
  url = {https://link.aps.org/doi/10.1103/PhysRevAccelBeams.27.060702}
}

@article{margraf_2023,
  author    = {Margraf, Rachel and Robles, River and Halavanau, Alex and Kryzywinski, Jacek and Li, Kenan and MacArthur, James and Osaka, Taito and Sakdinawat, Anne and Sato, Takahiro and Sun, Yanwen and Tamasaku, Kenji and Huang, Zhirong and Marcus, Gabriel and Zhu, Diling},
  title     = {Low-loss stable storage of 1.2 Å {X}-ray pulses in a 14 m {Bragg} cavity},
  journal   = {Nat. Photon.},
  year      = {2023},
  volume    = {17},
  number    = {10},
  pages     = {878--882},
  issn      = {1749-4893},
  doi       = {10.1038/s41566-023-01267-0},
  url       = {https://doi.org/10.1038/s41566-023-01267-0}
}

@article{benabderrahmane_prab_2017,
  title = {Development and operation of a {Pr}$_2${Fe}$_{14}${B} based cryogenic permanent magnet undulator for a high spatial resolution x-ray beam line},
  author = {C. Benabderrahmane and others},
  journal = {Physical Review Accelerators and Beams},
  volume = {20},
  issue = {3},
  pages = {033201},
  numpages = {14},
  year = {2017},
  month = {Mar},
  publisher = {American Physical Society},
  doi = {10.1103/PhysRevAccelBeams.20.033201},
  url = {https://link.aps.org/doi/10.1103/PhysRevAccelBeams.20.033201}
}

@article{bernhard_prab_2016,
  title = {Radiation emitted by transverse-gradient undulators},
  author = {Bernhard, Axel and Braun, Nils and Rodr\'{\i}guez, Ver\'onica Afonso and Peiffer, Peter and Rossmanith, Robert and Widmann, Christina and Scheer, Michael},
  journal = {Physical Review Accelerators and Beams},
  volume = {19},
  issue = {9},
  pages = {090704},
  numpages = {13},
  year = {2016},
  month = {Sep},
  publisher = {American Physical Society},
  doi = {10.1103/PhysRevAccelBeams.19.090704},
  url = {https://link.aps.org/doi/10.1103/PhysRevAccelBeams.19.090704}
}

@article{maiman_nature_1960,
  author    = {Maiman, T. H.},
  title     = {Stimulated Optical Radiation in Ruby},
  journal   = {Nature},
  year      = {1960},
  month     = aug,
  day       = {1},
  volume    = {187},
  number    = {4736},
  pages     = {493--494},
  issn      = {1476-4687},
  doi       = {10.1038/187493a0},
  url       = {https://doi.org/10.1038/187493a0}
}

@article{madey_japplphys_1971,
    author = {Madey, John M. J.},
    title = {{Stimulated Emission of Bremsstrahlung in a Periodic Magnetic Field}},
    journal = {Journal of  Applied Physics},
    volume = {42},
    number = {5},
    pages = {1906-1913},
    year = {1971},
    month = {04},
    issn = {0021-8979},
    doi = {10.1063/1.1660466},
    url = {https://doi.org/10.1063/1.1660466},
}

@article{agapov_nima_2015,
title = {Feasibility of a ring {FEL} at low emittance storage rings},
journal = {Nuclear Instruments and Methods in Physics Research A},
volume = {793},
pages = {35--40},
year = {2015},
issn = {0168-9002},
doi = {10.1016/j.nima.2015.04.069},
url = {https://doi.org/10.1016/j.nima.2015.04.069},
author = {I. Agapov}
}

@InProceedings{agapov_ipac_2018,
  author       = {I. V. Agapov and Y.-C. Chae and W. Hillert},
  title        = {{Low Gain FEL Oscillator Option for PETRA IV}},
  booktitle    = {Proc. 9th Int. Particle Accel. Conf. (IPAC'18)},
  pages        = {1420--1422},
  paper        = {TUPMF069},
  venue        = {Vancouver, BC, Canada},
  publisher    = {JACoW Publishing},
  address      = {Geneva, Switzerland},
  eventdate = {2018-04-29/2018-05-04},
  month        = {June},
  year         = {2018},
  isbn         = {978-3-95450-184-7},
  doi          = {10.18429/JACoW-IPAC2018-TUPMF069},
  url          = {http://jacow.org/ipac2018/papers/tupmf069.pdf},
}

@article{fisher_nima_1992,
title = {{40 Å FEL designs for the PEP storage ring}},
journal = {Nuclear Instruments and Methods in Physics Research A},
volume = {318},
number = {1},
pages = {730-735},
year = {1992},
issn = {0168-9002},
doi = {10.1016/0168-9002(92)91148-3},
url = {https://doi.org/10.1016/0168-9002(92)91148-3},
author = {Alan S. Fisher and Juan C. Gallardo and Heinz-Dieter Nuhn and Roman Tatchyn and Herman Winick and Claudio Pellegrini}
}

@article{jiao_aapps_2026,
  author    = {Yi Jiao and Ming Li and Ping He and Jingyi Li and Ye Tao and Gang Xu and Yuhui Dong and Weimin Pan},
  title     = {{Design and commissioning of the High Energy Photon Source}},
  journal   = {AAPPS Bull.},
  year      = {2026},
  volume    = {36},
  number    = {1},
  pages     = {3},
  issn      = {2309-4710},
  doi       = {10.1007/s43673-026-00184-y},
  url       = {https://doi.org/10.1007/s43673-026-00184-y},
  date      = {2026-03-02}
}

@article{shin_aapps_2021,
  author    = {Seunghwan Shin},
  title     = {New era of synchrotron radiation: fourth-generation storage ring},
  journal   = {AAPPS Bull.},
  year      = {2021},
  month     = aug,
  day       = {30},
  volume    = {31},
  number    = {1},
  pages     = {21},
  issn      = {2309-4710},
  doi       = {10.1007/s43673-021-00021-4},
  url       = {https://doi.org/10.1007/s43673-021-00021-4}
}

@article{casalbuoni_j_phys_conf_2025,
author = {Casalbuoni, S. and others},
doi = {10.1088/1742-6596/3094/1/012009},
url = {https://doi.org/10.1088/1742-6596/3094/1/012009},
year = {2025},
month = {sep},
publisher = {IOP Publishing},
volume = {3094},
number = {1},
pages = {012009},
title = {Status of {S-PRESSO}, a superconducting undulator for the {European XFEL}},
journal = {Journal of Physics: Conference Series}
}

@article{kesgin_ieee_2021,
  author={Kesgin, Ibrahim and others},
  journal={IEEE Transactions on Applied Superconductivity}, 
  title={{Fabrication and Testing of 18-mm-Period, 0.5-m-Long Nb$_3$Sn Superconducting Undulator}}, 
  year={2021},
  volume={31},
  number={5},
  pages={4100205},
  doi={10.1109/TASC.2021.3057846}
}

@article{huang_2012,
  title = {Compact X-ray Free-Electron Laser from a Laser-Plasma Accelerator Using a Transverse-Gradient Undulator},
  author = {Huang, Zhirong and Ding, Yuantao and Schroeder, Carl B.},
  journal = {Physical Review Letters},
  volume = {109},
  issue = {20},
  pages = {204801},
  numpages = {5},
  year = {2012},
  month = {Nov},
  publisher = {American Physical Society},
  doi = {10.1103/PhysRevLett.109.204801},
  url = {https://link.aps.org/doi/10.1103/PhysRevLett.109.204801}
}

@article{slusher_1999_rmp,
  title = {Laser technology},
  author = {Slusher, R. E.},
  journal = {Reviews of Modern Physics},
  volume = {71},
  issue = {2},
  pages = {S471--S479},
  numpages = {0},
  year = {1999},
  month = {Mar},
  publisher = {American Physical Society},
  doi = {10.1103/RevModPhys.71.S471},
  url = {https://link.aps.org/doi/10.1103/RevModPhys.71.S471}
}

@article{matthews_1995_nimb,
title = {Review of X-ray lasers},
journal = {Nuclear Instruments and Methods in Physics Research B},
volume = {98},
number = {1},
pages = {91-94},
year = {1995},
issn = {0168-583X},
doi = {10.1016/0168-583X(95)00079-8},
url = {https://doi.org/10.1016/0168-583X(95)00079-8},
author = {Dennis L. Matthews}
}

@article{suckewer_1990_science,
author = {S. Suckewer and C. H. Skinner},
title = {Soft X-Ray Lasers and Their Applications},
journal = {Science},
volume = {247},
number = {4950},
pages = {1553-1557},
year = {1990},
doi = {10.1126/science.2321016},
URL = {https://www.science.org/doi/abs/10.1126/science.2321016}}

@article{zhao_2010_rast,
author = {Zhao, Z. T.},
title = {{Storage Ring Light Sources}},
journal = {Reviews of Accelerator Science and Technology},
volume = {03},
number = {01},
pages = {57-76},
year = {2010},
doi = {10.1142/S1793626810000361}
}

@article{pellegrini_2020_natrevphys,
  author    = {Claudio Pellegrini},
  title     = {The development of {XFELs}},
  journal   = {Nature Revviews Physics},
  year      = {2020},
  month     = jul,
  volume    = {2},
  number    = {7},
  pages     = {330--331},
  issn      = {2522-5820},
  doi       = {10.1038/s42254-020-0197-1},
  url       = {https://doi.org/10.1038/s42254-020-0197-1}
}

@article{kim_2008_prl,
  title = {{A Proposal for an X-Ray Free-Electron Laser Oscillator with an Energy-Recovery Linac}},
  author = {Kim, Kwang-Je and Shvyd'ko, Yuri and Reiche, Sven},
  journal = {Physical Review Letters},
  volume = {100},
  issue = {24},
  pages = {244802},
  numpages = {4},
  year = {2008},
  month = {Jun},
  publisher = {American Physical Society},
  doi = {10.1103/PhysRevLett.100.244802},
  url = {https://link.aps.org/doi/10.1103/PhysRevLett.100.244802}
}

@article{wang_matradext_2020,
    author = {Wang, Shu-Xing and Zhu, Lin-Fan},
    title = {Non-resonant inelastic X-ray scattering spectroscopy: A momentum probe to detect the electronic structures of atoms and molecules},
    journal = {Matter and Radiatiation at Extremes},
    volume = {5},
    number = {5},
    pages = {054201},
    year = {2020},
    month = {07},
    issn = {2468-2047},
    doi = {10.1063/5.0011416},
    url = {https://doi.org/10.1063/5.0011416},
}

@misc{adams_workshop_2019,
      title={{Scientific Opportunities with an X-ray Free-Electron Laser Oscillator}}, 
    author={Bernhard Adams and others},
      year={2019},
      eprint={1903.09317},
      archivePrefix={arXiv},
      primaryClass={physics.ins-det},
      doi={10.48550/arXiv.1903.09317}, 
}

@article{barletta_nima_2010,
title = {{Free electron lasers: Present status and future challenges}},
journal = {Nuclear Instruments and Methods in Physics Research A},
volume = {618},
number = {1},
pages = {69-96},
year = {2010},
issn = {0168-9002},
doi = {https://doi.org/10.1016/j.nima.2010.02.274},
url = {https://www.sciencedirect.com/science/article/pii/S0168900210005656},
author = {W.A. Barletta and others}
}

@techreport{marie_2019,
  title = {Report of the High-Energy {X-ray} {FELs} Workshop: Exploring the New Frontier of High-Energy and High-Flux {X-ray} Free-Electron Lasers},
  address = {Santa Fe, New Mexico},
  month = {jul},
  year = {2019},
  url = {https://cdn.lanl.gov/files/document-45_1f184.pdf},
  institution = {Los Alamos National Laboratory}
}

@article{cai_2013_srn,
author = {Yunhai Cai and Yuantao Ding and Robert Hettel and Zhirong Huang and Lanfa Wang and Liling Xiao},
title = {An X-ray Free Electron Laser Driven by an Ultimate Storage Ring},
journal = {Synchrotron Radiatiation News},
volume = {26},
number = {3},
pages = {39--41},
year = {2013},
publisher = {Taylor \& Francis},
doi = {10.1080/08940886.2013.791216},
URL = {https://doi.org/10.1080/08940886.2013.791216}
}

@mastersthesis{hansen_2023_lund,
  author       = {Hansen, Jordis},
  title        = {Feasibility Study of {X}-ray Free Electron Lasers in Fourth Generation Storage Rings},
  school       = {Lund University, Department of Physics},
  year         = {2023},
  type         = {Bachelor's thesis},
  address      = {Lund, Sweden},
  url          = {http://lup.lub.lu.se/student-papers/record/9121693}
}

@article{colella_1984_optcommun,
title = {{Proposal for a free electron laser in the X-ray region}},
journal = {Optics Communications},
volume = {50},
number = {1},
pages = {41-44},
year = {1984},
issn = {0030-4018},
doi = {10.1016/0030-4018(84)90009-9},
url = {https://www.sciencedirect.com/science/article/pii/0030401884900099},
author = {R. Colella and A. Luccio}
}

@techreport{elegant,
  author       = {Borland, M},
  title        = {{ELEGANT}: A flexible {SDDS}-compliant code for accelerator simulation},
  institution  = {Argonne National Laboratory},
  doi          = {10.2172/761286},
  url          = {https://www.osti.gov/biblio/761286},
  address        = {Advanced Photon Source, Lemont, IL, USA},
  number = {LS-287},
  year         = {2000},
  month        = {08}}

@article{reiche_1999_nima,
title = {{GENESIS 1.3: a fully 3D time-dependent FEL simulation code}},
journal = {Nuclear Instruments and Methods in Physics Ressearch A},
volume = {429},
number = {1},
pages = {243--248},
year = {1999},
issn = {0168-9002},
doi = {10.1016/S0168-9002(99)00114-X},
url = {https://www.sciencedirect.com/science/article/pii/S016890029900114X},
author = {S. Reiche}
}

@unpublished{qin_ipac_2026a,
    author = {W. Qin and W. Liu and Y. Zhao and J. Wei and M. Li and X. Li and Y. Jiao and S. Wang},
    title = {{Feasibility of a Compact X-ray Free-Electron Laser Oscillator based on Diffraction Limited Storage Ring}},
    booktitle = {Proc. IPAC'26},
    paper = {TUP2606},
    venue = {Deauville, France},
    intype = {presented at IPAC'26},
    series = {IPAC'26 - 17th International Particle Accelerator Conference},
    number = {17},
    publisher = {JACoW Publishing, Geneva, Switzerland},
    month = {05},
    issn = {2673-5350},
    isbn = {978-3-95450-252-3},
    url = {https://jacow.org/ipac2026/pdf/TUP2606.pdf},
    note = {abstract submitted to IPAC'26, Deauville, France, 2026, paper TUP2606, May 2026 (unpublished)},
    eventdate = {2026-05-17/2026-05-22},
}

@article{first_tgu,
    author = {Smith, T. I. and Madey, J. M. J. and Elias, L. R. and Deacon, D. A. G.},
    title = {Reducing the sensitivity of a free‐electron laser to electron energy},
    journal = {Journal of Applied Physics},
    volume = {50},
    number = {7},
    pages = {4580-4583},
    year = {1979},
    month = {07},
    issn = {0021-8979},
    doi = {10.1063/1.326564},
    url = {https://doi.org/10.1063/1.326564}
}

@article{aneke_sri2024,
doi = {10.1088/1742-6596/3010/1/012032},
url = {https://dx.doi.org/10.1088/1742-6596/3010/1/012032},
year = {2025},
month = {may},
publisher = {IOP Publishing},
volume = {3010},
number = {1},
pages = {012032},
author = {Aneke, Emmanuel and Wootton, Kent P. and McChesney, Jessica and Zheng, Hao and Rodolakis, Fanny and Freeland, John},
title = {Simulation and Measurement of Horizontal Emittance via Undulator High Harmonics at the {APS-U}},
journal = {J of Physics: Conference Series}
}

@inproceedings{aneke_ipac2025,
    author = {E. Aneke and J. McChesney and K. Wootton and H. Zheng},
    title = {Measurement of vertical and horizontal emittance via undulator high harmonics at the {APS-U}},
    booktitle = {Proc. IPAC'25},
    pages = {2852-2855},
    paper = {THPM077},
    venue = {Taipei, Taiwan},
    series = {IPAC'25 - 16th International Particle Accelerator Conference},
    number = {16},
    publisher = {JACoW Publishing, Geneva, Switzerland},
    month = {06},
    year = {2025},
    issn = {2673-5490},
    isbn = {978-3-95450-248-6},
    doi = {10.18429/JACoW-IPAC2025-THPM077},
    url = {https://indico.jacow.org/event/81/contributions/8435},
    eventdate = {2025-06-01/2025-06-06},
}

@article{dattoli_1980,
  author  = {Dattoli, G. and Renieri, A.},
  title   = {Storage ring operation of the free-electron laser: The oscillator},
  journal = {Nuovo Cimento},
  volume  = {59},
  number  = {1},
  pages   = {1--39},
  year    = {1980},
  doi     = {10.1007/BF02739044},
  url     = {https://doi.org/10.1007/BF02739044},
  issn    = {1826-9877}
}

@article{fel1x,
  title = {Mode growth and competition in the x-ray free-electron laser oscillator start-up from noise},
  author = {Lindberg, R. R. and Kim, K.-J.},
  journal = {Physical Review Special Topics Accelerators and Beams},
  volume = {12},
  issue = {7},
  pages = {070702},
  numpages = {11},
  year = {2009},
  month = {Jul},
  publisher = {American Physical Society},
  doi = {10.1103/PhysRevSTAB.12.070702},
  url = {https://link.aps.org/doi/10.1103/PhysRevSTAB.12.070702}
}

@article{yuri_2010,
  author  = {Shvyd'ko, Yuri V. and Stoupin, Stanislav and Cunsolo, Alessandro and Said, Ayman H. and Huang, Xianrong},
  title   = {High-reflectivity high-resolution X-ray crystal optics with diamonds},
  journal = {Nature Physics},
  volume  = {6},
  number  = {3},
  pages   = {196--199},
  year    = {2010},
  doi     = {10.1038/nphys1506},
  url     = {https://doi.org/10.1038/nphys1506}
}

@article{elleaume_macro-temporal_1984,
	title = {Macro-temporal structure of storage ring free electron lasers},
	volume = {45},
	issn = {0302-0738},
	url = {http://www.edpsciences.org/10.1051/jphys:01984004506099700},
	doi = {10.1051/jphys:01984004506099700},
	number = {6},
	urldate = {2024-11-22},
	journal = {Journal de Physique},
	author = {Elleaume, P.},
	year = {1984},
	pages = {997--1001},
}

@InProceedings{dattoli_2019,
  author       = {G. Dattoli and M.-E. Couprie and C. Pellegrini},
  title        = {{Riding the FEL Instability (Dedicated to Alberto Renieri)}},
  booktitle    = {Proc. FEL'19},
  pages        = {1--6},
  paper        = {MOA01},
  venue        = {Hamburg, Germany},
  series       = {Free Electron Laser Conference},
  number       = {39},
  publisher    = {JACoW Publishing, Geneva, Switzerland},
  month        = {nov},
  year         = {2019},
  issn         = {""},
  isbn         = {978-3-95450-210-3},
  doi          = {10.18429/JACoW-FEL2019-MOA01},
  url          = {http://jacow.org/fel2019/papers/moa01.pdf},
}

@techreport{fornek_2019,
  author       = {Fornek, Thomas E.},
  title        = {Advanced Photon Source Upgrade Project Final Design Report},
  institution  = {Argonne National Laboratory (ANL), Argonne, IL (United States)},
  doi          = {10.2172/1543138},
  url          = {https://www.osti.gov/biblio/1543138},
  place        = {United States},
  year         = {2019},
  month        = {05}}

@article{goldstein_1985,
  title = {Theory of the sideband instability in free electron lasers},
  author = {Goldstein, J. C.},
  journal = {Nucl. Instrum. Methods Phys. Res. A},
  volume = {237},
  pages = {27},
  year = {1985},
  doi={10.1016/0168-9002(85)90325-0}
}

@article{colson_1986,
  title = {The trapped-particle instability in free electron laser oscillators and amplifiers},
  author = {Colson, W. B.},
  journal = {Nucl. Instrum. Methods Phys. Res. A},
  volume = {250},
  pages = {168--175},
  year = {1986},
  doi={10.1016/0168-9002(86)90878-8}
}

@InProceedings{raubenheimer_2018_FLS,
  author       = {T.O. Raubenheimer},
  title        = {{T}he {LCLS-II-HE}, {A} {H}igh {E}nergy {U}pgrade of the {LCLS-II}},
  booktitle    = {Proc. 60th ICFA Advanced Beam Dynamics Workshop (FLS'18),
                  Shanghai, China, 5-9 March 2018},
  pages        = {6--11},
  paper        = {MOP1WA02},
  venue        = {Shanghai, China},
  series       = {ICFA Advanced Beam Dynamics Workshop},
  number       = {60},
  publisher    = {JACoW Publishing},
  address      = {Geneva, Switzerland},
  month        = {June},
  year         = {2018},
  isbn         = {978-3-95450-206-6},
  doi          = {doi:10.18429/JACoW-FLS2018-MOP1WA02},
  url          = {http://jacow.org/fls2018/papers/mop1wa02.pdf},
  note         = {https://doi.org/10.18429/JACoW-FLS2018-MOP1WA02},
}

@misc{suppMaterial,
  note = {See Supplemental Material at [URL will be inserted by publisher] for `Feasibility of Insertion Devices', which includes Refs. \cite{casalbuoni_j_phys_conf_2025, kesgin_ieee_2021, bernhard_prab_2016, benabderrahmane_prab_2017}.}
}

@misc{suppMaterial2,
  note = {See Supplemental Material at [URL will be inserted by publisher] for `Bragg-crystal Optics Model', which includes Ref. \cite{yuri_2010}.}
}

@misc{suppMaterial3,
  note = {See Supplemental Material at [URL will be inserted by publisher] for `A Single Insertion Device'.}
}

@misc{tanaka_2026,
  author = {Hitoshi Tanaka},
  title = {{Synergies between FELs and ring-based SR sources}},
  howpublished = {presented at Temporally Coherent Synchrotron Light Sources, Hamburg, Germany},
  year = {2026},
  month = {jan},
  url = {https://indico.desy.de/event/51344/contributions/195466/}
}

\end{document}



\preprint{SRFELO paper 1}

\title{Realizing A Hard X-Ray Storage Ring Free Electron Laser Oscillator at the APS-U -- Supplemental Material}

\author{Emmanuel Aneke}
\email{eaneke@u.northwestern.edu}
\affiliation{Applied Physics Program, Northwestern University, Evanston, Illinois 60208, USA}

\author{Chris Jacobsen}
\email{c-jacobsen@northwestern.edu}
\affiliation{Department of Physics \&{} Astronomy, Northwestern University, Evanston, Illinois 60208, USA}

\author{Ryan Lindberg}
\email[Corresponding author: ]{lindberg@anl.gov}
\affiliation{Argonne National Laboratory, 9700 S Cass Avenue, Lemont, Illinois 60439, USA}

\author{Kent P. Wootton}
\email[Corresponding author: ]{kwootton@anl.gov}
\affiliation{Argonne National Laboratory, 9700 S Cass Avenue, Lemont, Illinois 60439, USA}


\date{\today}


\maketitle



\section{Feasibility of Insertion Devices}

The planar insertion device needed for the 5~keV oscillator is feasible, at a period of $\lambda_u=1.94$~cm and gap of 6~mm ($K=2.25$). We consider this well within the state-of-the-art demonstrated by different technologies including superconducting undulators based on both NbTi \cite{casalbuoni_j_phys_conf_2025} and Nb$_3$Sn \cite{kesgin_ieee_2021}. To support the 8~keV and 10~keV oscillators, a TGU would be employed. Superconducting TGUs have been experimentally demonstrated \cite{bernhard_prab_2016}. For the 8~keV case, a TGU is needed at $K_0 = 1.73$ and a period of $\lambda_u=1.7$~cm. A cryogenic permanent magnetic undulator with very similar performance has been demonstrated at SOLEIL \cite{benabderrahmane_prab_2017}. The 10~keV case employs the same undulator period at a smaller deflection parameter $K_0=1.42$.

\section{Bragg-crystal Optics Model}
The cavity requires wavelength-selective crystal Bragg reflectors for near-normal incidence reflectivity \cite{yuri_2010} and CRLs for transverse mode stability. To overcome the inherent absorptive and scattering losses of these elements, we found that the FEL interaction must provide a single-pass gain explicitly exceeding ${\sim}$5\% which, in turn, guided our outcoupling threshold for all cases. Furthermore, the optical round-trip time must precisely match the electron bunch arrival interval. In this initial study we have considered filling with 48 bunches: this does not preclude operating the ring with more flexible modes. By filling the storage ring with a specific harmonic pattern (48 bunches), the trapped radiation pulse amplifies by interacting with a fresh electron bunch on each consecutive round trip, preventing sequential degradation of a single bunch. The cavity length was selected to be 23.99\,m. Lastly, to maximize the transverse spatial overlap and optimize the fundamental 3D FEL coupling, the optical cavity was configured such that the Rayleigh ranges of the radiation mode were explicitly matched to the betatron functions of the electron beam.

Accurately representing these narrow-band crystal optics is essential for evaluating the oscillator's spectral dynamics and multi-pass stability. In our tracking framework, the Bragg mirrors are modeled using the dynamical theory of X-ray diffraction. For a given target photon energy, we generate complex reflectivity profiles containing both the real and imaginary components of the crystal response, uniquely determined by the chosen Miller indices of the diamond lattice. Crucially, the penetration of the X-ray field into the crystal lattice introduces a wavelength-dependent phase shift and longitudinal group delay governed by the Fresnel equations. To computationally isolate the FEL gain dynamics from this optical delay, we introduced a cavity length correction of length $\ell$ in Fourier space by multiplying by $e^{i\ell\omega/c}$. This effectively ``flattens'' the phase response of the generated Bragg mirror profiles. This ``phase flattening'' ensures that the simulated optical wavefront remains perfectly synchronous with the reference macroparticle. Thus mimicking the real-world procedure of physically detuning the absolute cavity length by a few microns to account for the propagation delay inside the crystals.

\section{A Single Insertion Device}
In a physical TGU, the relative transverse magnetic gradient $\alpha$ is primarily a geometric constant dictated by the undulator period and the fixed cant angle of the magnetic poles; it does not scale with the baseline gap distance. Consequently, when the gap is opened to transition from $K=1.73$ (8.05~keV) to $K=1.42$ (10~keV), the fundamental TGU energy-spread cancellation condition ($\alpha D = [2+K^2]/K^2$) is maintained not by altering the magnetic gradient, but by dynamically tuning the incident lattice dispersion $D$. By utilizing the storage ring's upstream lattice optics to proportionally increase the dispersion, the requisite resonance cancellation is perfectly preserved across the energy tuning range. Furthermore, any corresponding shifts in the dimensionless TGU scaling parameter $\Gamma$ can be compensated for by localized tuning of the betatron function $\beta_y$, ensuring the optimal spatial overlap between the electron beam and the cavity mode remains fully intact. However, increasing $\beta_y$ has the consequence of increasing the beam size which decreases the peak electron density and, in turn, reduces the gain. Ultimately, the emittance coupling can be adjusted to compensate for this gain degradation.

\bibliography{bibtex/references.bib}